\documentclass[pra,superscriptaddress,twocolumn,amsmath,amssymb]{revtex4}
\usepackage{amsfonts}
\usepackage{amssymb}
\usepackage{graphicx}
\usepackage{subfigure}
\usepackage{color}
\usepackage[normalem]{ulem}
\usepackage{natbib}
\usepackage{bbold}
\usepackage{placeins}
\usepackage{soul,xcolor}
\usepackage{epstopdf}
\usepackage{tabu}
\usepackage{array}
\usepackage{upgreek}
\usepackage{braket}
\usepackage{tikz}
\DeclareMathAlphabet{\mathpzc}{OT1}{pzc}{m}{it}
\usepackage{xcolor}

\usepackage{multirow}
\usepackage[utf8]{inputenc}
\usepackage[colorlinks = truelinkcolor = blue, urlcolor  = blue, citecolor = blue, anchorcolor = blue]{hyperref}
\hypersetup{
	colorlinks   = true, 
	urlcolor     = blue, 
	linkcolor    = blue, 
	citecolor   = blue 
}

\begin{document}

	\title{ Quantum Zeno effect and nonclassicality in a PT symmetric system of coupled cavities }

	\author{Javid Naikoo}
	\thanks{naikoo.1@iitj.ac.in}
	\affiliation{Indian Institute of Technology Jodhpur, Jodhpur 342011, India}
	\author{Kishore Thapliyal}
	\thanks{tkishore36@yahoo.com}
	\affiliation{Jaypee Institute of Information Technology, A-10, Sector-62, Noida UP-201307, India}
	\affiliation{RCPTM, Joint Laboratory of Optics of Palacky University and Institute of Physics of Academy of Science of the Czech Republic, Faculty of Science, Palacky University, 17. listopadu 12, 771 46 Olomouc, Czech Republic}
	\author{Subhashish Banerjee}
	\thanks{subhashish@iitj.ac.in}
	\affiliation{Indian Institute of Technology Jodhpur, Jodhpur 342011, India}
    \author{Anirban Pathak}
    \thanks{anirban.pathak@jiit.ac.in}
    \affiliation{Jaypee Institute of Information Technology, A-10, Sector-62, Noida UP-201307, India}

	\date{\today} 
	
	\begin{abstract}
     The interplay between  the nonclassical features and the parity-time (PT) symmetry (or its breaking) is studied here by considering  a PT symmetric system consisting of two cavities with gain and loss. The conditions for PT invariance is obtained for this system.  The behavior of the average photon number corresponding to the gain and loss modes for  different initial states (e.g., vacuum, NOON,  coherent, and thermal states) has also been obtained. With the help of the number operators, quantum Zeno and anti-Zeno effects are studied, and the observed behavior is compared in PT symmetric (PTS) and PT symmetry broken (PTSB) regimes. It has been observed that the relative phase of the input coherent fields plays a key role in the occurrence of these effects. Further, some nonclassicality features are witnessed using criteria based on the number operator(s). Specifically, intermodal antibunching, sum and difference squeezing, are investigated for specific input states. It is found that the various nonclassical features, including the observed quantum Zeno and anti-Zeno effects, are suppressed when one goes  from  PTS to  PTSB regime. In other words, the dominance of the loss/gain rate in the field modes over the coupling strength between them diminishes the nonclassical features of the system.
	\end{abstract}

	\maketitle 
	
	\section{Introduction}
	  The quantum systems are in many ways different from their classical counterparts. The most fundamental distinction is in the way they respond to  a measurement or an interaction. The interaction of a quantum system with the measuring device has profound consequences on its subsequent dynamics, and can even suppress the time evolution if the interaction is frequent enough, a phenomenon known as the  quantum Zeno effect (QZE) \cite{misra1977zeno}. The QZE has been recently realized in many experiments and its applcations have been reported  in quantum information \cite{barenco1997stabilization}, avoiding the decoherence \cite{beige2000quantum, berman2000suppression, zhou2009quantum}, to sustain the entanglement \cite{maniscalco2008protecting, wang2008quantum}, in the purification of quantum systems \cite{erez2008thermodynamic}, to suppress the intermolecular forces \cite{wuster2017quantum}, and to realize direct counterfactual communication \cite{cao2017direct}. The converse phenomenon of  QZE is referred to as quantum anti-Zeno effect (QAZE) in which the time evolution of the quantum system speeds up when the measurements are frequent enough. The QZE and QAZE have been observed in many systems. For example, in trapped ions and atoms  \cite{itano1990quantum, fischer2001observation}, superconducting qubits \cite{barone2004dynamical,harrington2017quantum,kakuyanagi2015observation}, Bose-Einstein condensates \cite{streed2006continuous}, nanomechanical oscillators \cite{chen2010quantum}, quantum cavity systems \cite{helmer2009quantum} and nuclear spin systems \cite{wolters2013quantum, zheng2013experimental, kalb2016experimental}. In \cite{segal2007zeno, chaudhry2016general, eleuch2017gain, zhou2017quantum, chaudhry2014zeno, zhang2015zeno, maniscalco2006zeno, PhysRevA.98.012135}, the QZE and QAZE have been studied in the context of open quantum systems, too. Another interesting feature of QZE that has been studied in the recent times is the formulation of a joint strategy by two or more players leading to their emerging as winners, and is broadly referred to as quantum Parrondo's game \cite{PhysRevLett.83.3077,DTratchet,meyer2002quantum,chandrashekar2011parrondo}.
	  \par The quantum Zeno effect is just one nontrivial consequence of the interaction between two quantum systems. There are many more. For example, the  interactions between two systems can also lead to the inseparability of their quantum states, entanglement. Various optical/optomechanical systems have also been designed to generate the desired nonclassical states \cite{pathak2018classical} of radiation, thereby bringing the quantum aspects in the table top experiments. Different facets of nonclassicality, characterized by the negative values of Glaubler-Sudarshan $P$ function \cite{glauber1963coherent,sudarshan1963equivalence}, have been extensively investigated in various systems. A set of single mode nonclassical features (\cite{agarwal2013quantum} and references therein), like sub-Poissonian photon statistics, antibunching, and squeezing of a field, have been reported to be useful in the development of quantum inspired technology \cite{dowling2003quantum,browne2017quantum}. Two field modes may show nonlocal correlations as entanglement \cite{horodecki2009quantum}, steering \cite{cavalcanti2016quantum}, and Bell nonlocality \cite{brunner2014bell} having applications in secure quantum communication \cite{gisin2002quantum,pathak2018quantum}.  Various  witnesses of quantumness, including the ones mentioned here, have been studied in many systems, viz.,  cavity and optical systems \cite{sen2013intermodal, Alam2017, naikoo2017probing, thapliyal2014higher,thapliyal2014nonclassical}, Bose-Einstein condensates \cite{giri2014single, Giri2017}, optomechanical systems \cite{Alam2017, bose1997preparation}, atoms and quantum dots \cite{baghshahi2015generation, majumdar2012probing}, single and interacting qubits \cite{thapliyal2015quasi, chakrabarty2010study}, and engineered quantum states \cite{thapliyal2017comparison,malpani2018lower}.

	Contemporary to the development of quantum optics, has been the  emergence of parity-time (PT) symmetric optics, where the notion of PT symmetry is introduced to explain the real spectrum of non-Hermitian Hamiltonians \cite{bender1998real, bender2007making}.  The interest in this phenomenon has been escalated in the recent times \cite{PhysRevLett.100.103904, PhysRevLett.101.080402, PhysRevLett.104.054102, PhysRevLett.103.093902, Ruter, PhysRevLett.106.093902, Regensburger, PT-whisprngGalry, PhysRevA.90.033804,liu2017controllable}.  The PT symmetric (PTS) Hamiltonian ($H$) can have a real eigenvalue spectrum despite being non-Hermitian \cite{bender1998real}. Specifically,  $[H, PT]=0$ assures the real eigenvalue spectrum of $H$. For example, $\hat{p}^2 + i \hat{x}^3$ and $\hat{p}^2 - \hat{x}^4$ are not Hermitian but PTS and possess real eigenvalues. In fact, these two Hamiltonians are special cases of the general parametric family of PTS Hamiltonians $H = \hat{p}^2 + \hat{x}^2 (i \hat{x})^{\epsilon}$, such that for $\epsilon \ge 0$  all the  eigenvalues are real while for $\epsilon <0$ they are complex. These two regimes are respectively known as PTS and PT symmetry broken (PTSB) regimes \cite{bender2007making}.  An equivalence of a quantum system possessing PT symmetry and a quantum system having Hermitian Hamiltonian was shown in \cite{mostafazadeh2003exact}.  In \cite{PT-whisprngGalry}, a system was realized whose dynamics is governed by PT Hamiltonian. Many optomechanical properties have been investigated  for PTS systems, such as the cavity optomechanical properties underlying the phonon lasing action \cite{PTPhononLaser},  PTS chaos \cite{PTchaos}, cooling of mechanical oscillator \cite{liu2016energy}, cavity assisted metrology \cite{liu2016metrology}, optomechanically-induced-transparency \cite{PT-OIT}, and optomechanically induced absorption  \cite{PT-OIA}. The possibility of the spontaneous generation of photons in PTS systems is illustrated in \cite{Agarwal2}. In \cite{Agarwal1}, the gain in the quantum amplification by the superradiant emission of radiation was shown to be a consequence of the broken PT symmetry. Further, the exceptional points for an optical coupler with one lossy waveguide and polarization entangled input states were obtained in \cite{Longhi:18}. Nonclassicality in the coherent states for non-Hermitian systems is also reviewed recently \cite{dey2018squeezed}.\par
	In this work, we aim to study the behavior of the various nonclassical features of a system as one  goes from  PTS to PTSB regime. We analyze the effect of this transition on the possibility of presence of QZE and QAZE as well as the nonclassical features, such as intermodal antibunching and the sum and difference squeezing for different choices of the input states. The rest of the paper is planned as follows. In Sec. \ref{modelsec}, we discuss the model and the solution to  the equations of motion of cavity field modes in the Heisenberg picture. Section \ref{analysis} is devoted to the discussion  of various nonclassical features of the field modes. We finally conclude in Sec. \ref{conclusion}.

	\begin{figure}[ht] 
		\centering
		\begin{tabular}{cc}
			\includegraphics[width=60mm]{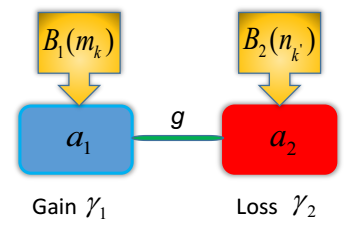}
		\end{tabular}
		\caption{(Color online) The model. Two cavities  bearing modes $a_1$ and $a_2$ coupled through coupling constant $g$ are also interacting with baths $B_1$ and $B_2$, respectively. The baths cause gain $\gamma_1$ and loss $\gamma_2$ in the first and second cavity, respectively.}
		\label{model}
	\end{figure}

	\section{Model and Solution} \label{modelsec}
	
\begin{figure}[ht] 
		\centering
		\begin{tabular}{cc}
			(a)\includegraphics[width=70mm]{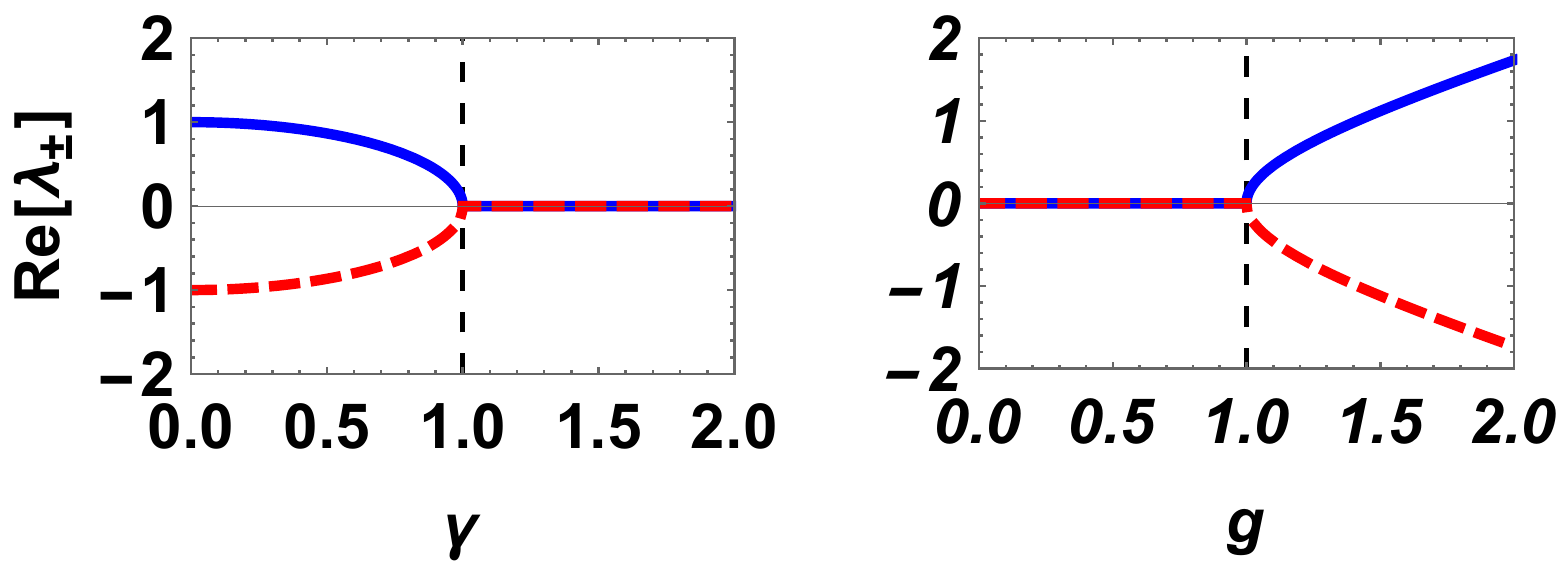}\\
			(b)\includegraphics[width=70mm]{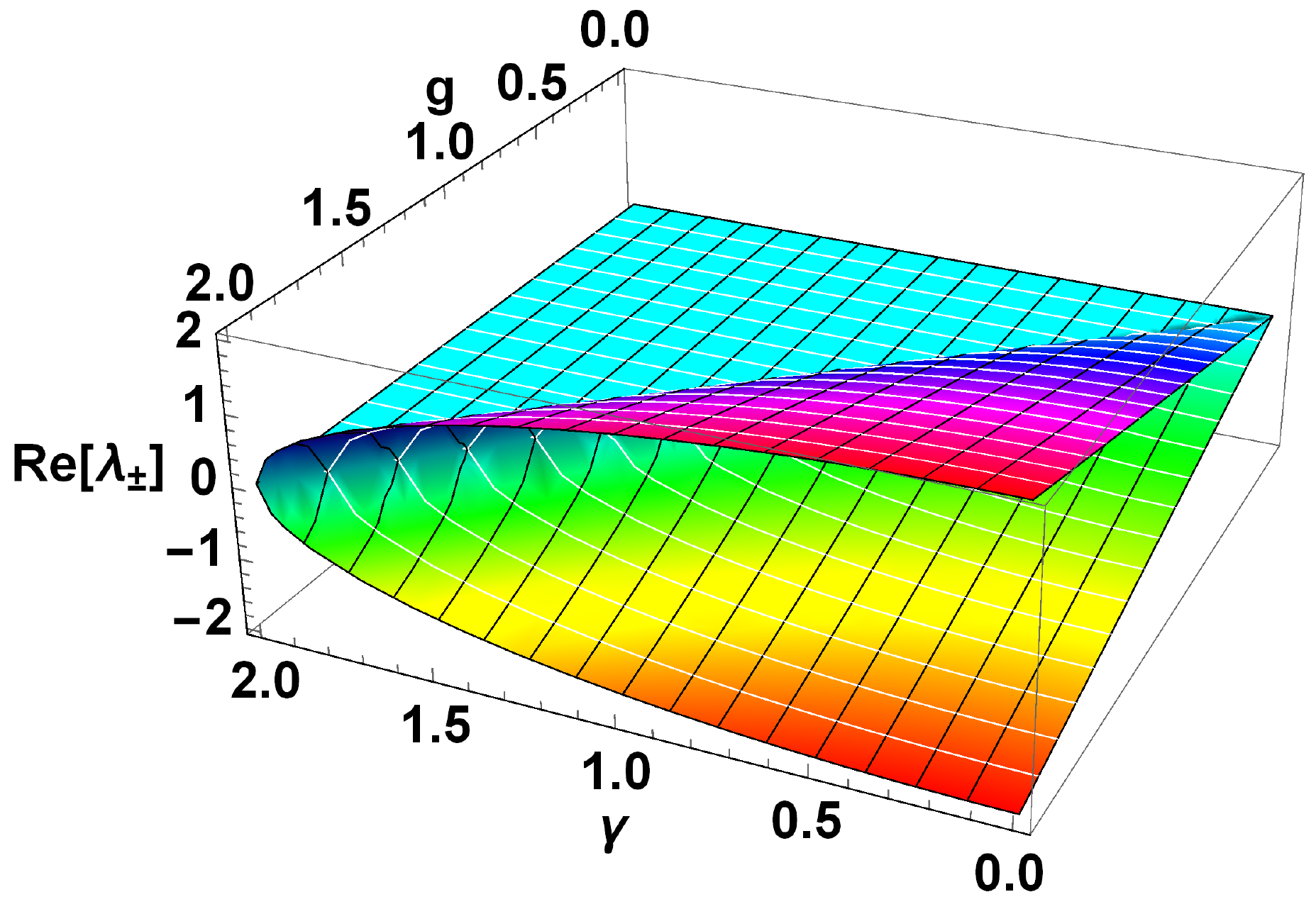}
		\end{tabular}
		\caption{(Color online) The real part of the eigenvalues $\lambda_{\pm}$ is plotted as a function of the coupling strength $g$ and the gain (loss) rate $\gamma$. The points where the two eigenvalues coalesce are called as exceptional points. In (a), the blue (solid) and red (dashed) curves correspond to $\lambda_+$ and $\lambda_{-}$, respectively.}
		\label{eigenvalues}
	\end{figure}

	 In this work, we are interested to study the interplay between PT symmetry and various facets of nonclassicality. To this effect, we consider the system sketched in Fig. \ref{model}.  Two optical cavities bearing modes $a_1$ and $a_2$, with corresponding frequencies $\omega_1$ and $\omega_2$, are connected by coupling constant $g$. The Hamiltonian for this system can be written as
	 \begin{align}\label{HS}
	H_S &= \omega_1 a_1^\dagger  a_1 +  \omega_2 a_2^\dagger  a_2  + g (a_1^\dagger a_2 + {\rm H.c.}),
      \end{align}
where H.c. stands for Hermitian conjugate. Throughout this paper, we are going to work in the natural units ($\hbar = c = 1$). To bring the $PT$ symmetric effects, we allow  the cavities in the system of interest to interact with the ambient environmental degrees of freedom. We denote the baths (reservoirs) as $B_1$ and $B_2$ and consider them to be coupled to the cavities bearing modes $a_1$ and $a_2$, respectively. We further assume that the former cavity has a gain rate $\gamma_1$, and the later has a loss rate $\gamma_2$. The Hamiltonian pertaining to the baths $(H_B)$ and the system-bath interaction $(H_{SB})$ are respectively  given by  
	\begin{subequations} 
	\begin{eqnarray}
	H_B &=& \sum_{k} \nu_k m_k^\dagger m_k + \sum_{k^\prime} \nu_{k^\prime}  n_{k^\prime}^\dagger n_{k^\prime}, \label{HB}\\
	H_{SB} &=& \left\{\sum_{k} g_k m_k^\dagger a_1  + \sum_{k^\prime} g_{k^\prime} n_{k^\prime}^\dagger a_2  + {\rm H.c.}\right\}.\label{HSB}
	\end{eqnarray}
\end{subequations}
	Here, $m_k$ and $n_k^\prime$ are the annihilation operators corresponding to the baths $B_1$ and $B_2$, respectively, and are coupled to the corresponding cavity modes $a_1$ and $a_2$ with coupling strengths $g_k$ and $g_{k^\prime}$. Using Eqs. (\ref{HS}), (\ref{HB}), and (\ref{HSB}), we obtain the following Langevin equations: \begin{subequations} 
	\begin{eqnarray}
	\dot{a}_1(t) &= -i \omega_1 a_1(t) + \gamma_1 a_1(t) + f_1(t) - i g a_2(t),\label{a1dot}\\
	\dot{a}_2(t) &= -i \omega_2 a_2(t) - \gamma_2 a_2(t) + f_2(t) - i g a_1(t). \label{a2dot}
	\end{eqnarray}
\end{subequations}
	Here, $f_1(t)$ and $f_2(t)$ are the noise operators given by $-i \sum_{l} g_l b_l(0) e^{-i\nu_l t}$, where $b_l(0)$ denotes the corresponding bath operator. The noise operators satisfy the following properties \cite{agarwal2013quantum}: 
	\begin{subequations} 
	\begin{eqnarray}
	\langle f_1^\dagger(t) f_1(t^\prime) \rangle &= 2 \gamma_1 \delta(t-t^\prime), \quad \langle f_1(t) f_1^\dagger(t^\prime) \rangle = 0,\label{f1prop}\\
	\langle f_2(t) f_2^\dagger(t^\prime) \rangle &= 2 \gamma_2 \delta(t-t^\prime), \quad \langle f_2^\dagger(t) f_2(t^\prime) \rangle = 0.\label{f2prop}
	\end{eqnarray}
    \end{subequations}
	The action of the parity operator $P$ and the time reversal operator $T$ on modes $a_1$ and $a_2$ can be summarized as  \begin{subequations} 
	\begin{eqnarray}
	P&: a_1 \leftrightarrow -a_2, \quad a_1^\dagger \leftrightarrow - a_2^\dagger, \\
	T&: a_1 \leftrightarrow a_1, \quad a_1^\dagger \leftrightarrow a_1^\dagger, \quad a_2 \leftrightarrow a_2, \quad a_2^\dagger \leftrightarrow a_2^\dagger.
	\end{eqnarray}
\end{subequations}
	The action of the time reversal operator also flips the sign of the complex number $i$. Thus, the PT invariance of Eqs. (\ref{a1dot}) and (\ref{a2dot}) demands that
	\begin{equation}\label{PTconds}
         \omega_1 = \omega_2 = \omega \quad    {\textrm {and}}    \quad    \gamma_1 = \gamma_2 = \gamma.
	\end{equation}
	  To investigate the PT invariance further, let us redefine the annihilation operators as $\tilde{a}_1(t) = e^{-i\omega t} a_1(t)$, $\tilde{a}_2(t) = e^{-i\omega t} a_2(t)$, and  the noise operator $F_i(t) = e^{-i\omega t} f_i(t)$. With this transformation Eqs. (\ref{a1dot}) and (\ref{a2dot}) become 
	  \begin{subequations} 
	\begin{eqnarray}
	  \dot{\tilde{a}}_1(t) &=  \gamma \tilde{a}_1(t) + F_1(t) - i g \tilde{a}_2(t),\label{a1tildedot}\\
	  \dot{\tilde{a}}_2(t) &=   -\gamma \tilde{a}_2(t) + F_2(t) - i g \tilde{a}_1(t). \label{a2tildedot}
	\end{eqnarray}
\end{subequations}
	  One can write the formal solution of the above equations as follows	  
	\begin{equation}\label{sols}
	\begin{pmatrix}
      \tilde{a}_1(t)\\
      \tilde{a}_2(t)  
	\end{pmatrix}  = e^{-i \mathcal{K} t} \begin{pmatrix}
	                           \tilde{a}_1(0)\\
	                           \tilde{a}_2(0)
	                         \end{pmatrix}  + \int_{0}^{t}  ds \, e^{-i\mathcal{K} (t-s)} \begin{pmatrix}
	                                                                     F_1(s)\\
	                                                                     F_2(s)
	                                                               \end{pmatrix}.
	\end{equation}
	Here, $\mathcal{K}$ is identified as the effective Hamiltonian for the system given by 
	\begin{equation}\label{K}
	\mathcal{K} = \begin{pmatrix}
	                i\gamma &  g\\
	                g    & -i \gamma 
	              \end{pmatrix}
	\end{equation} 
	with eigenvalues 
	\begin{equation}\label{evsHeff}
	\lambda_{\pm} = \begin{cases}
	   \pm \sqrt{g^2-\gamma^2}  &  {\rm for~} g \geq \gamma ,\\\\
	   \pm i \sqrt{\gamma^2 - g^2}  &  {\rm for~}  g < \gamma. \\
	                \end{cases} 
	\end{equation}
	Apart from the conditions given in Eq. (\ref{PTconds}), the complete PT invariance demands that the eigenvalues of $\mathcal{K}$ are real, that is $\gamma \leq g$. Naturally, PTSB regime is characterized by $\gamma > g$. In other words, the dominance of the gain/loss over the coupling strength breaks the PT symmetry of the system. The transition from the PTS to PTSB regime is governed by the eigenvalues of the effective Hamiltonian. Figure \ref{eigenvalues} shows the behavior of the eigenvalues with respect to the coupling strength $g$ and the gain (loss) rate $\gamma$.  The two real branches of eigenvalues coalesce at $g = \gamma$ and become complex for $g < \gamma$. These points at which the transition from real to complex spectrum occurs, are known as \textit{exceptional points} \cite{Longhi:18}.

	We can now rewrite the solution given in Eq. (\ref{sols}) by setting $Q=e^{-i\mathcal{K} t}$. With $\mathcal{K}$ given in Eq. (\ref{K}), it can be shown that
	\begin{equation}\label{Qmat}
	Q = \begin{pmatrix}
	     \cosh(\Omega t) + \frac{ \gamma }{\Omega} \sinh (\Omega t) &  \frac{-ig}{\Omega}  \sinh(\Omega t)\\\\
	      \frac{ -ig }{\Omega} \sinh(\Omega t) & \cosh(\Omega t) -  \frac{\gamma }{\Omega} \sinh (\Omega t)	      
	    \end{pmatrix}.
	\end{equation}
	Here, $\Omega = \sqrt{\gamma^2 - g^2}$ controls the transition from PTS to PTSB phase. Finally, the solution turns out to be 
\begin{align}
\tilde{a}_1(t) &= Q_{11}(t) \tilde{a}_1(0) + Q_{12}(t) \tilde{a}_2(0) \nonumber \\&+ \int_{0}^{t} ds \bigg( Q_{11}(t-s) F_{1}(s) + Q_{12}(t-s) F_{2}(s) \bigg), \label{a1tilde} \\
\tilde{a}_2(t) &= Q_{21}(t) \tilde{a}_1(0) + Q_{22}(t) \tilde{a}_2(0) \nonumber \\& +\int_{0}^{t} ds \bigg( Q_{21}(t-s) F_{1}(s) + Q_{22}(t-s) F_{2}(s) \bigg). \label{a2tilde}
\end{align}
	One can obtain the solution at the exceptional points by taking  appropriate limits, specifically considering   $\Omega \rightarrow 0$, we can obtain
	\begin{equation}
	Q|_{\Omega \rightarrow 0} = \begin{pmatrix}
	                                1 + \gamma t  & -ig t\\
	                                 -ig t        & 1 - \gamma t    
	                            \end{pmatrix}.
	\end{equation}
Having obtained the solution for the two field modes $\tilde{a}_1(t)$ and $\tilde{a}_2(t)$, we now proceed to study some  properties of the output fields, like average photon numbers with different input states, and also look for the nonclassical features of the fields. Since the phase factor in $\tilde{a}_k(t) = e^{-i\omega t} a_k (t)$ ($k=1,2$) is not relevant in our study, in what follows, we would drop the tilde. \par

\section{Some properties of the  output fields} \label{analysis}
In this section, we analyze some properties associated with the field modes $a_1$ (gain) and $a_2$ (loss), and their behavior in PTS and PTSB regimes.\par
\textit{Average photon number:} We begin this study with the average photon number $n_{a_i} = \langle a_i^\dagger (t) a_i(t) \rangle$ corresponding to the mode $a_i$ $(i=1,2)$, by choosing different initial states. For example, with the input state as vacuum, one can obtain the following closed form expressions for the average photon number:
\small
\begin{align}
n_{a_1} &= \frac{2 \gamma^2 \Omega \cosh(2\Omega t)  - (\Omega^2 + \gamma^2) \gamma \sinh (2 \Omega t) - 2\gamma \Omega (\gamma + g^2 t) }{2 \Omega^3},\nonumber\\
n_{a_2} &= \frac{g^2 \gamma}{2 \Omega^2} \Big[ -2t  + \frac{\sinh (2 \Omega t)}{\Omega}  \Big].\label{num-vac}
\end{align}
\normalsize

\begin{figure} 
	\centering
	\includegraphics[width=85mm,height=150mm]{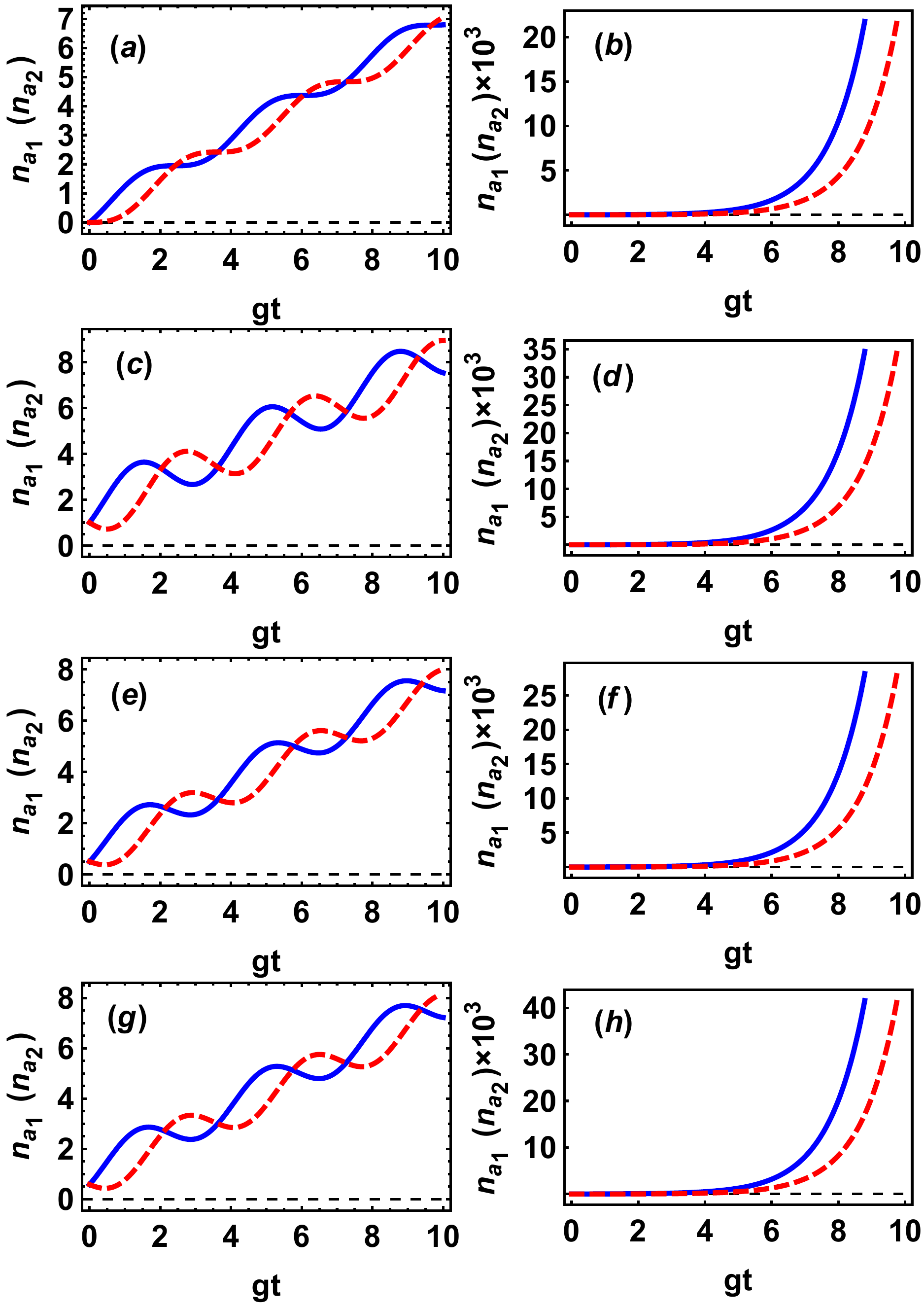}
	\caption{(Color online) Average photon number $n_{a_1} = \langle a_1^\dagger(t) a_1(t) \rangle$ (solid blue curve) and $n_{a_2} = \langle a_2^\dagger(t) a_2(t) \rangle$ (dashed red curve) with respect to the dimensionless parameter $gt$ for PTS (left panel) and PTSB (right panel) cases. The value of $\gamma$ is $0.5g$ and $1.1g$ corresponding to the PTS and PTSB regimes, respectively. The input states are: (a)-(b) Vacuum state $\ket{0 0}$; (c)-(d) Coherent state $\ket{\alpha_1 \alpha_2}$, with $\alpha_k=r_k e^{i \theta_k}$ for $k=1,2$ and coherent state parameters $r_1=r_2=1$,  $\theta_1=\theta_2=\pi/4$; (e)-(f) NOON state $(\ket{10} + \ket{01})/\sqrt{2}$; (g)-(h) Thermal state $\rho_0= (1-e^\beta)^2 \exp[-\beta(a_1^\dagger a_1 + a_2^\dagger a_2)]$ with $\beta = \hbar \omega/kT$. Here, we have chosen $\beta = 1$.}
	\label{AvgNumComb}
\end{figure}

Similarly, we have considered different initial states, such as coherent state $\ket{\alpha_1, \alpha_2}$, NOON state $(\ket{1,0} + \ket{0,1})/\sqrt{2}$, and thermal state $\rho_0= (1-e^\beta)^2 \exp[-\beta(a_1^\dagger a_1 + a_2^\dagger a_2)]$ (see Appendix {\bf A}), to compute the average photon numbers in the two cavities. The average photon number in each case is plotted in Fig. \ref{AvgNumComb}.  The parameters $\gamma$ (gain/loss rate) and $g$ (coupling strength) are chosen such that the system is either in PTS or PTSB regime. In the PTS regime, the average photon number for the gain and loss modes is observed to grow together, waning  the distinction between gain and loss cavities. In PTSB regime, however, the average photon number in the gain cavity grows  faster as compared to the average photon number in the  lossy cavity. This  is due to the fact that in the PTSB phase, the gain/loss dominates the coupling strength between the two cavities.  The oscillatory behavior of the curves in PTS case can be attributed to the fact that the elements of the $Q$ matrix change from hyperbolic to sinusoidal function as one goes from PTSB to PTS regime. The rapid increase in the photon number as a spontaneous photon generation process in the context of PT symmetry   was also reported in \cite{Agarwal2} in a system of two coupled waveguides.  In all these cases, {in PTS regime,} one can clearly see initial decay in the average photon number in the lossy cavity, which is compensated later by its interaction with the gain medium. In the set of possible input states, we have considered vacuum (shown to play an important role in PTS property \cite{Agarwal2}), a quantum state with positive (coherent state) and negative (NOON state) Glauber-Sudarshan $P$ function, and a mixed (thermal) state having positive $P$ function. Average photon numbers of two cavities does not give any signature of quantumness. Therefore, in what follows, we investigate the QZE and QAZE and some nonclassical features, like intermodal antibunching and squeezing, in the field modes, which will use the number operators calculated so far.

\textit{Quantum Zeno and anti-Zeno effects:} A more general definition of QZE involves the dynamics for which the interaction part may be defined as a `continuous gaze' on the system under consideration (see \cite{facchi2001quantum} for a review). This interaction may be a measurement operator to explain QZE as introduced in \cite{misra1977zeno}.  In the present case, the two cavity model (in Fig. \ref{model}) can be considered as a \textit{system-probe} configuration, where one of the cavities (considered system) is under a constant influence of the other cavity (probe). The occurrence of QZE and QAZE in the system-probe setting can be studied by defining Zeno parameter, introduced in \cite{KTzenoConf,KishoreZeno},
\begin{align}\label{zenodef}
\zeta_{a_i}(t) &= \frac{ n_{a_i} -n_{a_i}|_{g=0}}{\displaystyle \prod_{i=1,2} n_{a_i}},
\end{align}
with $n_{a_i} = \langle a_i^\dagger (t) a_i(t) \rangle$. Here, we have normalized the Zeno parameter by dividing by the product of the average number of photons of the two modes.
A positive (negative) value of the Zeno parameter $\zeta_{a_i}$ implies an increase (decrease) in the average photon numbers corresponding to the mode $a_i$ as a consequence of the coupling ($g$) with the probe field. The scenarios $\zeta_{a_i}(t) < 0$ and $\zeta_{a_i}(t) > 0$ are respectively known as QZE and QAZE.

\begin{figure}[t] 
	\centering
	(a)\includegraphics[width=70mm]{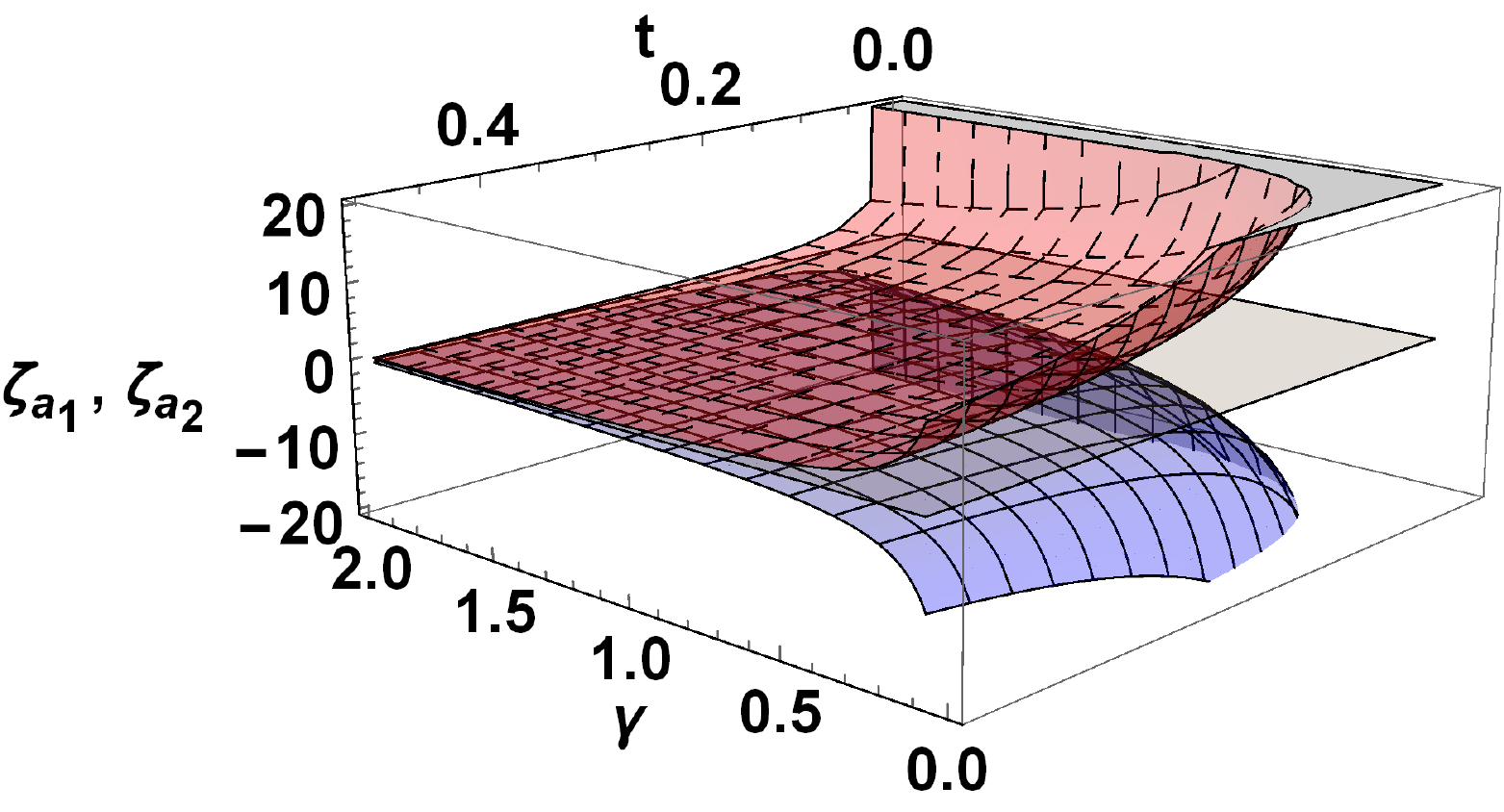}
	(b)\includegraphics[width=70mm]{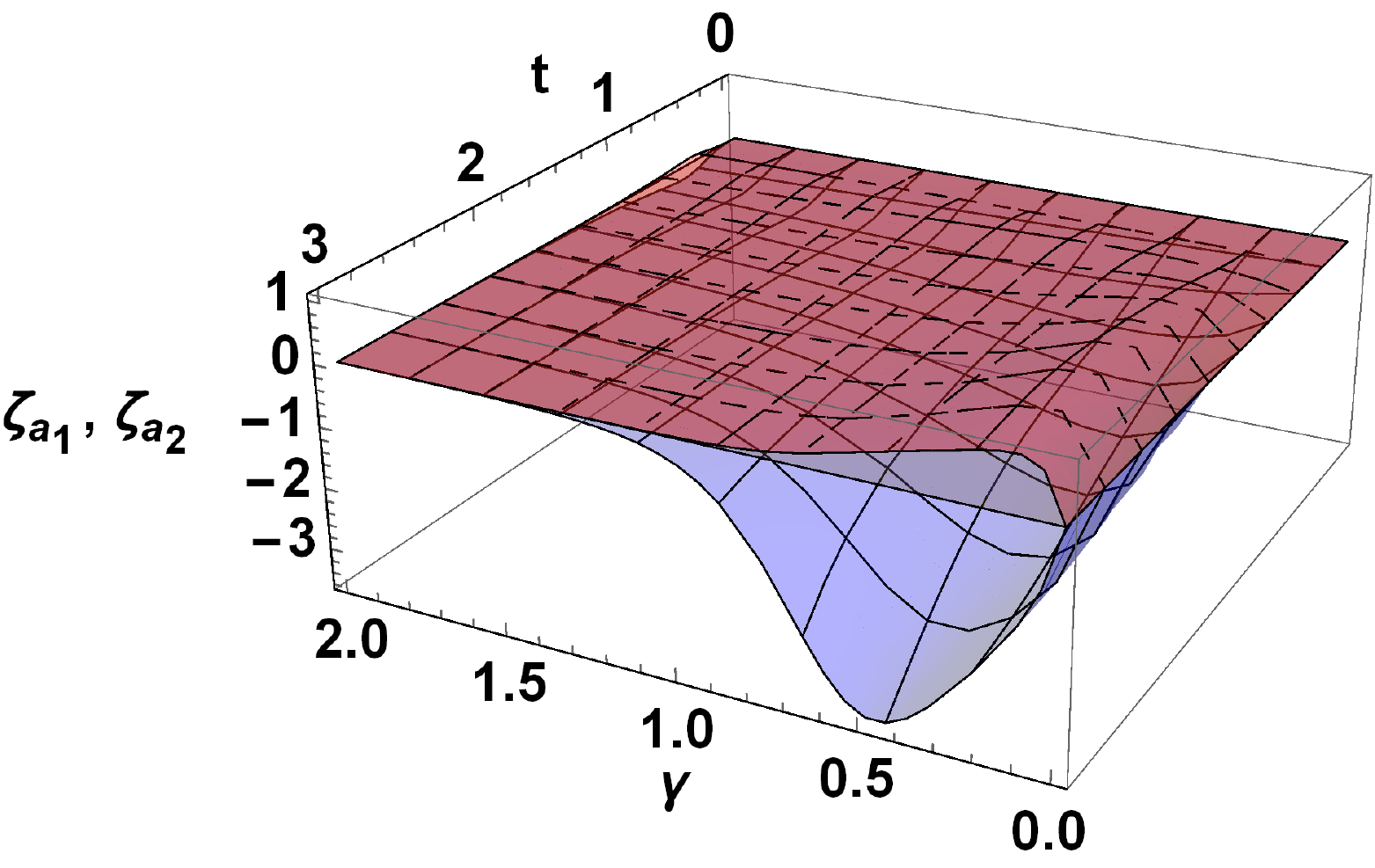}
	(c)\includegraphics[width=70mm]{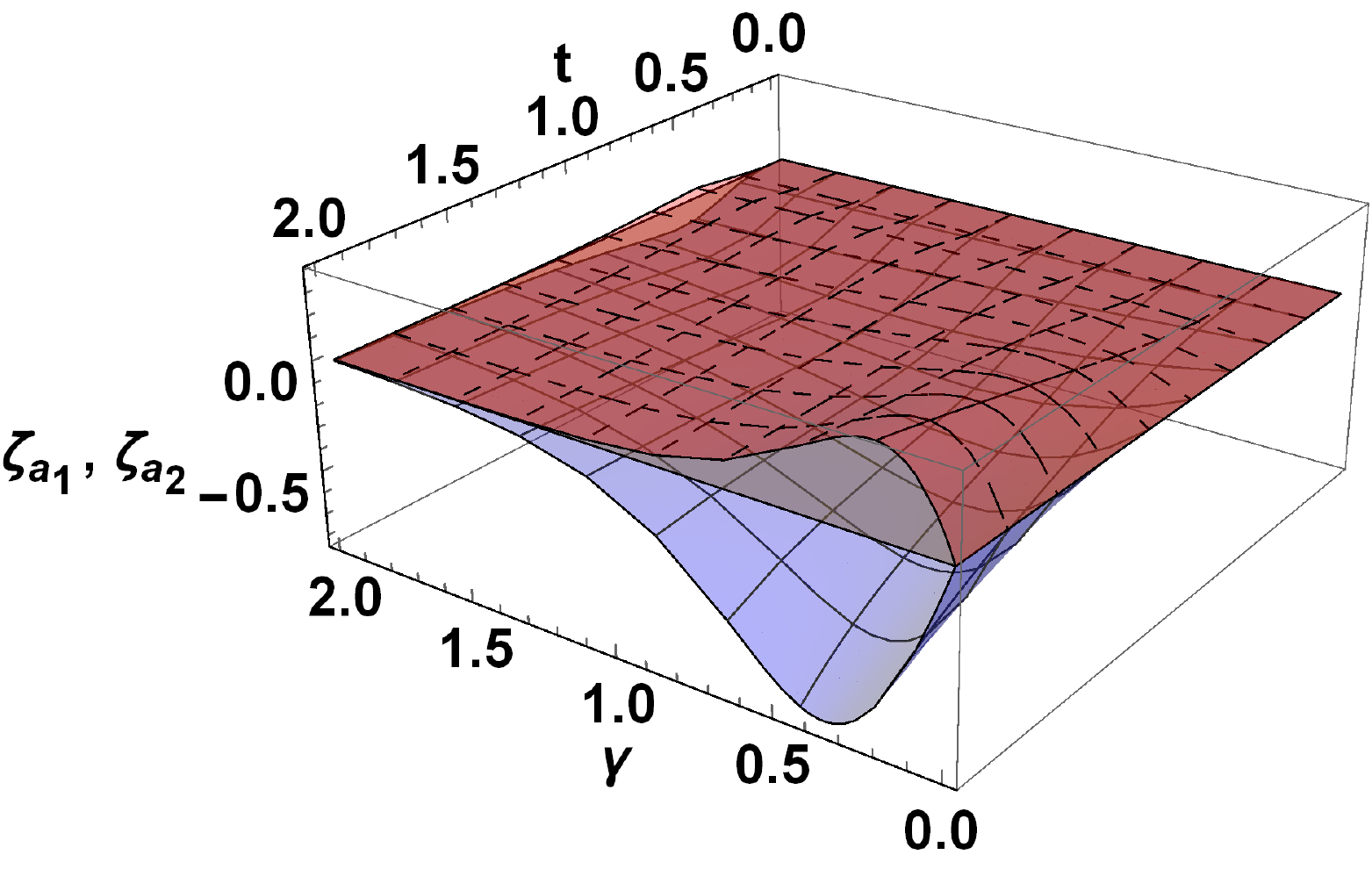}
	\caption{(Color online) Zeno parameter as defined in Eq. (\ref{zenodef}), $\zeta_{a_1}$ (blue surface) and $\zeta_{a_2}$ (red surface) with input state as vacuum (a), NOON state $(\ket{10} + \ket{01})/\sqrt{2}$ (b), and thermal state (c). In all the cases, the lossy mode ($a_2$) shows the QAZE while the gain mode ($a_1$) shows the QZE. Here, we have chosen coupling strength $g=1$, so that $\gamma <1$ and $\gamma > 1$ correspond to PTS and PTSB regimes, respectively.}
	\label{zeno_noon_thr}
\end{figure}

 Figure \ref{zeno_noon_thr} depicts the Zeno parameter with different initial states, viz., vacuum state (a), NOON state $(\ket{10} + \ket{01})/\sqrt{2}$ (b), and thermal state (c).  In all the cases, mode $a_2$  (red surface) shows the QAZE effect while  QZE is displayed by mode $a_1$ (blue surface). {This nature is observed due to the fact that the number of photons generated under an independent evolution of the gain cavity is suppressed (which is described as QZE) due to its interaction with the lossy cavity. In contrast, an increase in the number of photons (which is described as QAZE) in the lossy cavity is the outcome of its interaction with the gain cavity. This increase/decrease in the number of photons also depends upon the values of parameters deciding PT symmetry property of the system.}

\begin{figure}[t] 
	\centering
	(a)\includegraphics[width=80mm]{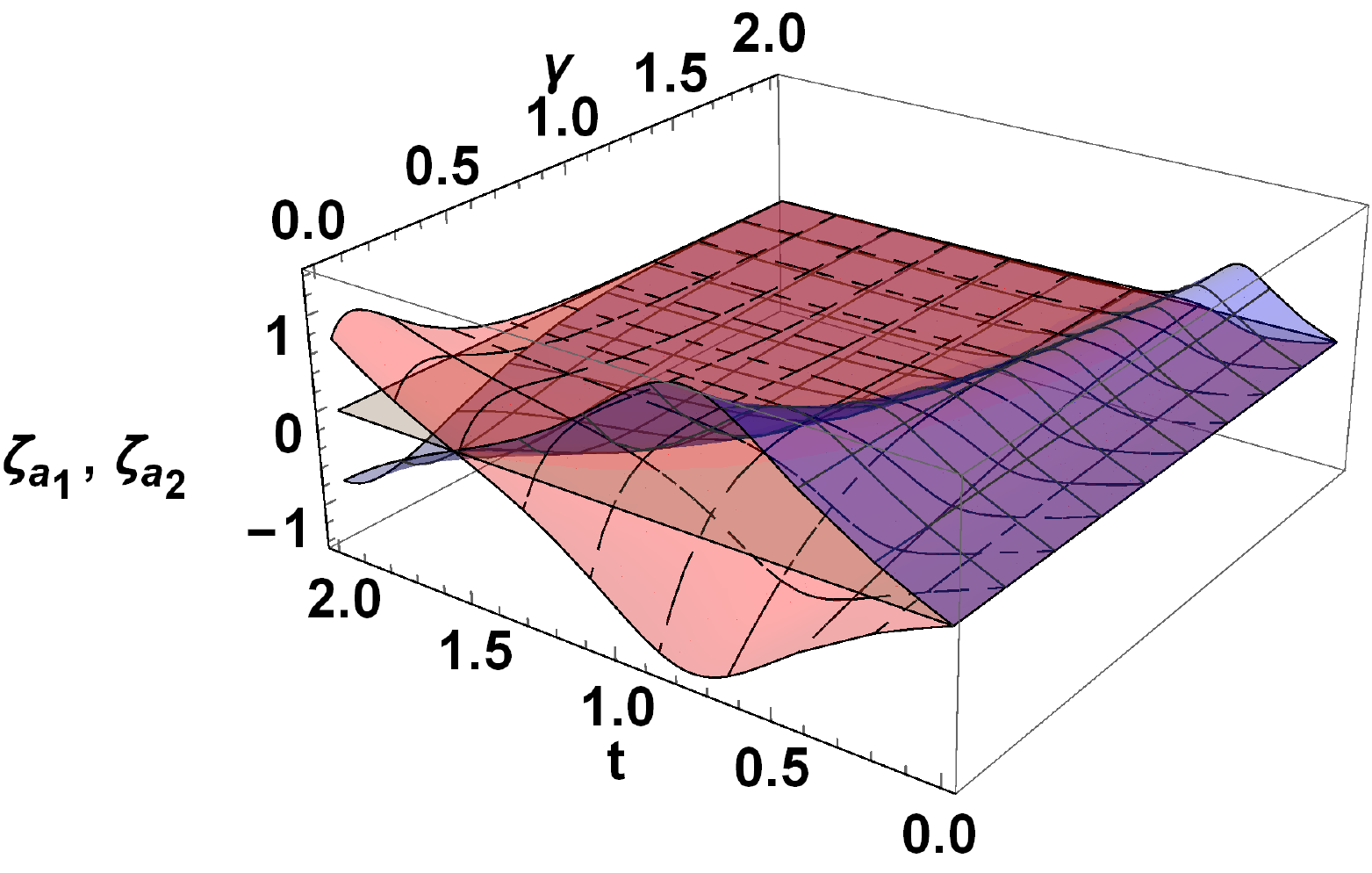}\\
	(b)\includegraphics[width=80mm]{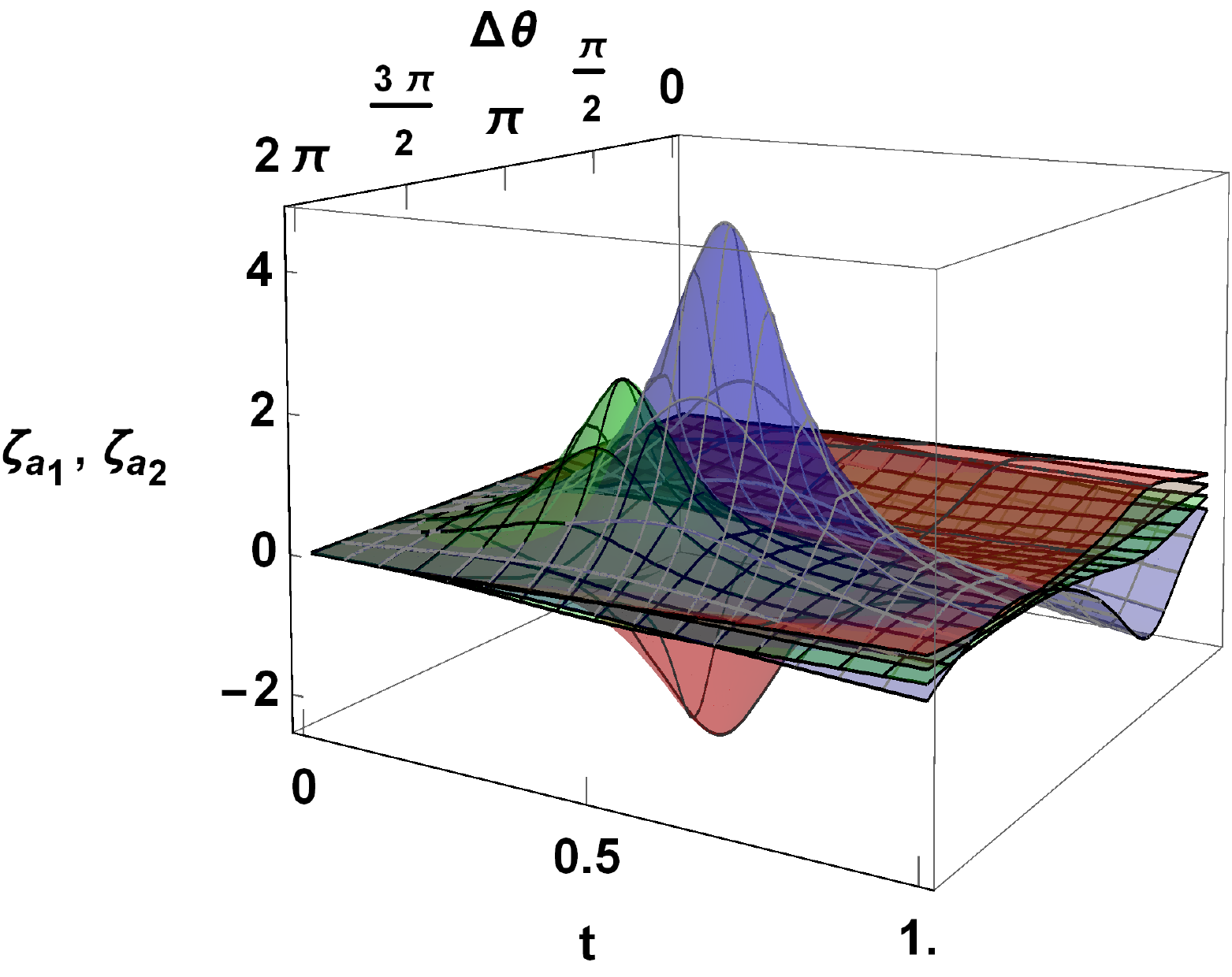}
	\caption{(Color online) Zeno parameter as defined in Eq. (\ref{zenodef}) with input state as coherent state $\ket{r_1 e^{i \theta_1}, r_2 e^{i\theta_2}}$ for $r_1=r_2=1$. In (a), $\theta_1= \pi$, $\theta_2 = -\pi/4$. The blue and red surfaces correspond to $\zeta_{a_1}$ and $\zeta_{a_2}$, respectively. Here, the  coupling strength between the cavities $g=1$.  (b) Variation with respect to the relative phase parameter $\Delta \theta = \theta_1 - \theta_2$.  The color scheme is as follows: blue  for $\zeta_{a_1}$, red  for $\zeta_{a_2}$ with $\gamma=0.5g$, that is, PTS regime;  green  for $\zeta_{a_1}$ and gray for $\zeta_{a_2}$ with $\gamma=1.5g$, PTSB regime. The parameter $\Delta \theta$ decides which of the two modes ($a_1$ or $a_2$) would show the QZE/QAZE. The maxima and minima in the plot occur  at $\Delta \theta = \pi/2, 3\pi /2$.}
	\label{zeno_coh}
\end{figure}

 We separately discuss the case when both the cavity fields are initially in the coherent states as in this case, the transition between the QZE and QAZE can be controlled by the parameters of the input fields.  Specifically, Fig. \ref{zeno_coh} depicts the Zeno parameter corresponding to modes $a_1$ and $a_2$ with input state as the coherent state $\ket{\alpha_1 \alpha_2}$, such that $\alpha_k = r_k e^{i\theta_k}$ with $k=1,2$. Figure \ref{zeno_coh} (a) shows the variation of the Zeno parameters with respect to the gain/loss rate $\gamma$ and time $t$. In PTS regime ($\gamma <g$), the QZE and QAZE  are more prominent as compared to PTSB regime ($\gamma > g$). { The observed behavior can be attributed to the fact that in the PTS phase, the coupling is dominant and has pronounced effect, i.e., losses in cavity mode $a_2$ are supplemented by the gain cavity due to strong coupling between them. This causes large variation in the Zeno parameter in PTS phase when compared with the PTSB phase.} In Fig. \ref{zeno_coh} (b),  the Zeno parameter is shown as a function of the relative phase (difference of the phases corresponding to the coherent states of the two modes) $\Delta \theta  = \theta_1 - \theta_2$ and time $t$. It is clear that the presence of QZE or QAZE in modes $a_1$ and $a_2$ depends on the value of $\Delta \theta$. In this case, for $ \Delta \theta > \pi $, the mode $a_1$ dominantly shows QAZE, while  for $\Delta \theta  < \pi$ it shows QZE.   {Therefore, a transition between QZE and QAZE can be controlled by the relative phase of the input coherent states, while variation in the amount of the Zeno parameter also depends upon whether the system is in the PTS/PTSB phase.}

\begin{figure}[t] 
	\centering
	(a)\includegraphics[width=80mm]{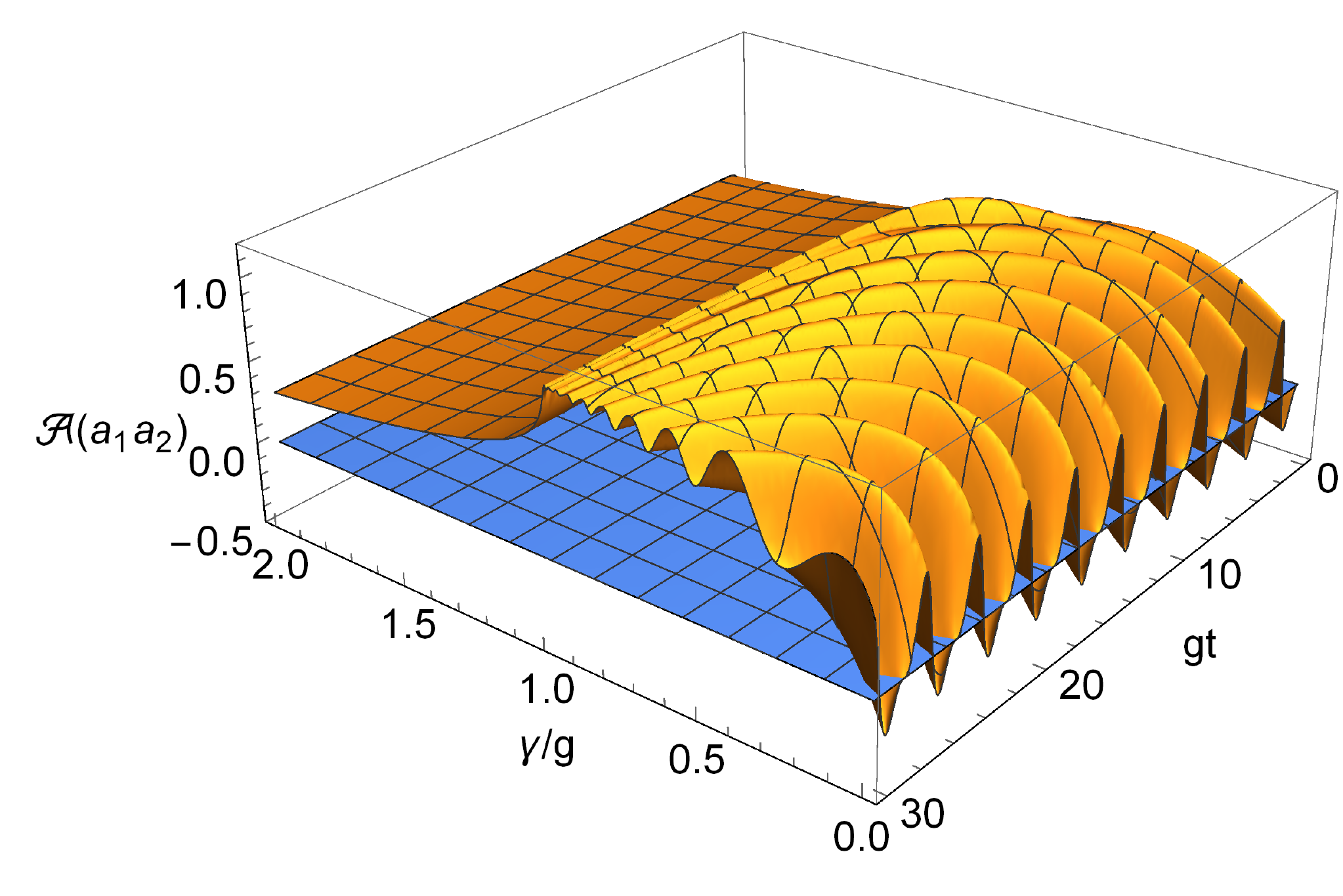}\\
	(b)\includegraphics[width=80mm]{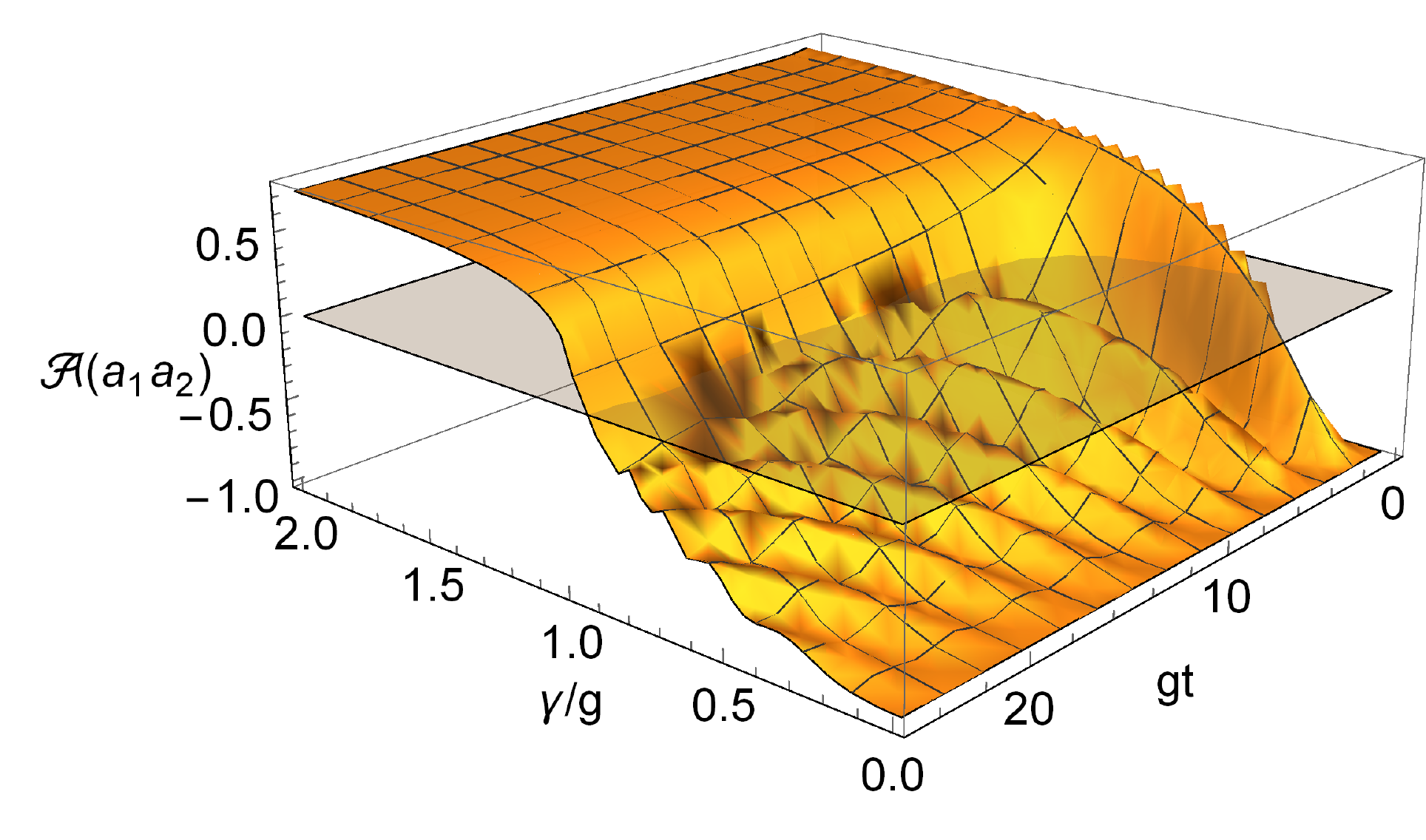}
	\caption{(Color online) Intermodal antibunching with input state as coherent state $\ket{\alpha_1, \alpha_2}$ (a) and NOON state $(\ket{10}+\ket{01})/\sqrt{2}$ (b). In the former $\alpha_1 = r_1 e^{i\theta_1}$ and $\alpha_2 = r_2 e^{i\theta_2}$ with $r_1=1$, $r_2=2$, $\theta_1=\theta_2=\pi/2$. The nonclassical behavior corresponds to $\mathpzc{A} (a_1 a_2) < 0$. The behavior in PTS regime ($\gamma < g$) is very different from the PTSB regime ($\gamma > g$).}
	\label{InterAntiBun}
\end{figure}

\textit{Intermodal antibunching:} For the field modes $a_1$ and $a_2$, the condition for intermodal antibunching is given as follows
\begin{equation}
\mathpzc{A} (a_1 a_2) = \langle a^\dagger_1 a_2^\dagger a_1 a_2 \rangle - \langle a_1^\dagger a_1 \rangle \langle a_2^\dagger a_2 \rangle < 0. 
\end{equation}
The first term in the right-hand side corresponds to the simultaneous detection in the outputs of two cavities, while the second term represents the product of individual detections in the outputs. 
In order to compute the first expectation value, we make use of the following decoupling relation (\cite{naikoo2017probing} and references therein)
\begin{align}
\langle A B C D \rangle & \approx \langle AB \rangle \langle CD \rangle + \langle AD \rangle \langle BC \rangle + \langle AC \rangle \langle BD \rangle \nonumber \\&- 2\langle A \rangle \langle B \rangle \langle C \rangle \langle D \rangle.
\end{align}
Thus, we obtain the average value of the witness of intermodal antibunching $\mathpzc{A} (a_1 a_2)$ for different initial states, which detects the presense of nonclassicality for the negative values of the witness $\mathpzc{A} (a_1 a_2)$. Figure \ref{InterAntiBun} depicts the variation of the intermodal antibunching witness $\mathpzc{A} (a_1 a_2)$ with input state as (a) coherent state and (b) NOON state. The nonclassical features are observed in both the cases as depicted by the negative values of the witness. Further, it is clear that the behavior in PTS and PTSB regimes is remarkably different, revealing that PTS phase favors nonclassicality compared to PTSB phase.\par

\begin{figure}[t] 
	\centering
	(a)\includegraphics[width=70mm]{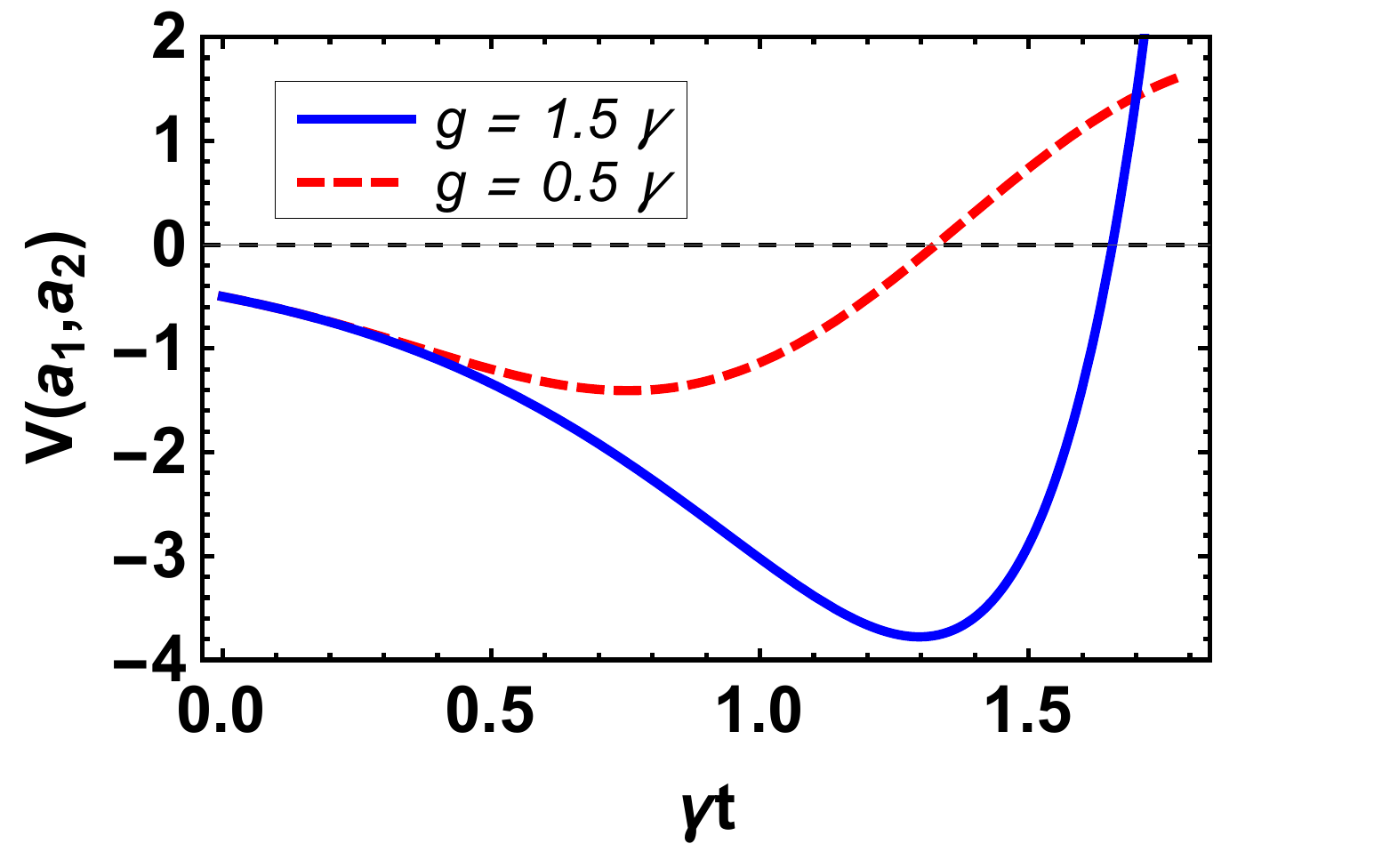}\\
	(b)\includegraphics[width=70mm]{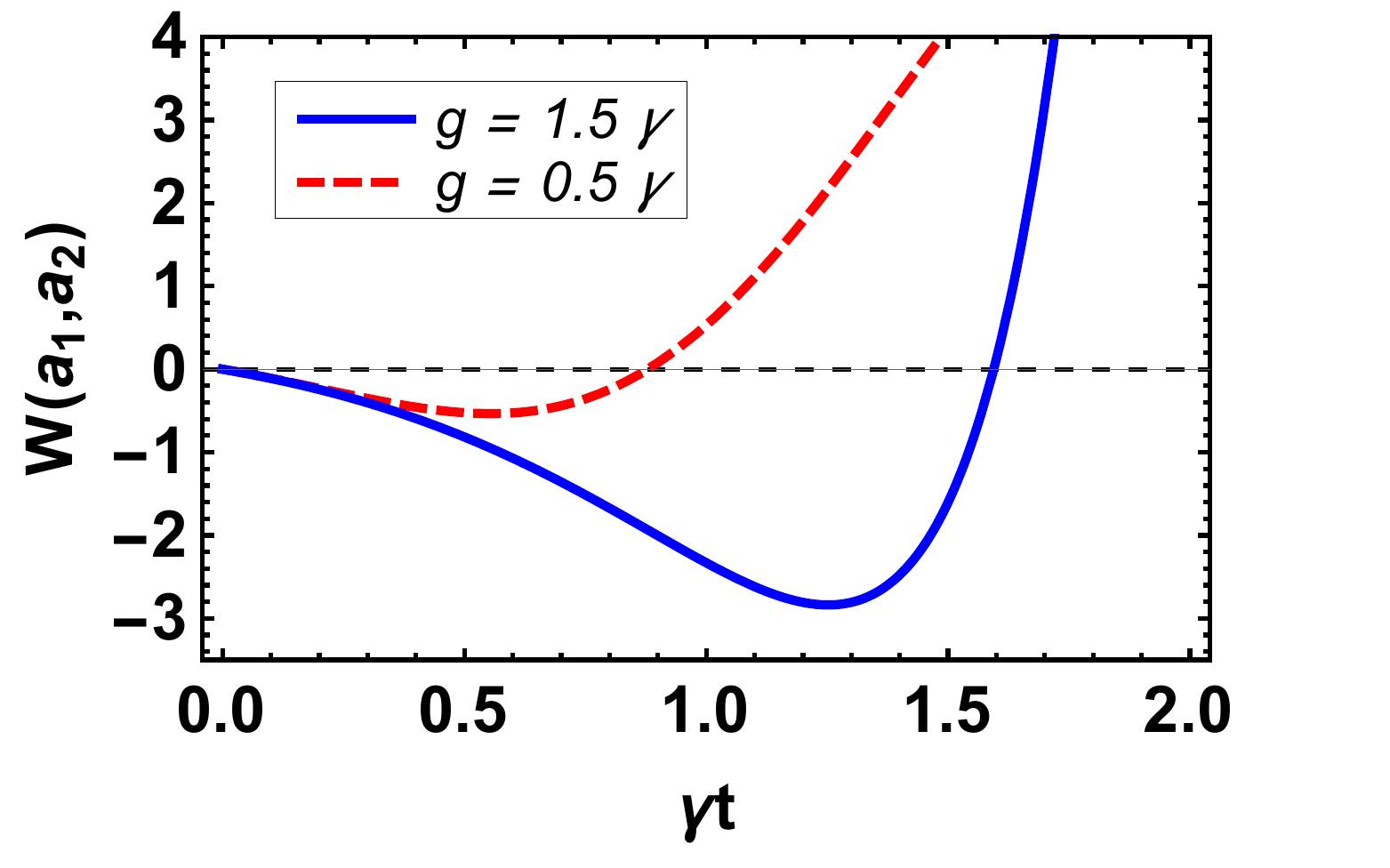}
	\caption{(Color online) Sum squeezing parameter $V(a_1,a_2)$ (a) and difference squeezing parameter $W(a_1, a_2)$ (b),  as defined in Eqs. (\ref{sumsqeezdef}) and (\ref{diffsqueezdef}), plotted against dimensionless parameter $\gamma t$  with vacuum as the initial state. A state is sum (difference) squeezed  if $V(a_1,a_2) <0$ $(W(a_1,a_2)<0)$. Here, we used $\phi = \pi/4$.}
	\label{sumsqueez}
\end{figure}

\textit{Sum squeezing criterion:} Hillery's sum squeezing criterion \cite{SumSqueezHillery} is defined in terms of a generalized two mode quadrature operator of the form 
\begin{equation}
V_{\phi} = \frac{e^{-i\phi} a_1 a_2 + e^{i\phi} a_1^\dagger a_2^\dagger }{2}, 
\end{equation} 
in analogy of the single mode quadrature
where $\phi$ is the phase angle of the coherent field used in the homodyne measurement. A state is said to be sum squeezed along phase angle $\phi$, if 
\begin{equation}\label{sumsqeezdef}
V(a_1, a_2) = \langle (\Delta V_\phi)^2 \rangle - \frac{ \langle a_1^\dagger a_1 \rangle + \langle a_2^\dagger a_2 \rangle + 1 }{4} < 0
\end{equation}
with $\langle (\Delta V_\phi)^2 \rangle = \langle V_\phi^2 \rangle - \langle V_\phi \rangle^2$. \par
\textit{Difference squeezing criterion:} A state is said to be difference squeezed if 
\begin{equation}\label{diffsqueezdef}
W(a_1, a_2) = \langle (\Delta W_\phi)^2 \rangle - \frac{ |\langle a_1^\dagger a_1 \rangle - \langle a_2^\dagger a_2 \rangle |}{4} < 0.
\end{equation}
The collective operator $W_\phi =\frac{1}{2} (e^{i\phi} a_1 a_2^\dagger + e^{-i\phi} a_1^\dagger a_2)$ and variance $\langle (\Delta W_\phi)^2 \rangle = \langle W_\phi^2 \rangle - \langle W_\phi \rangle^2$. 

We have chosen to study the sum and difference squeezing here as these two-mode nonclassical features use the average photon numbers we have studied in the beginning of this section. Figure \ref{sumsqueez} depicts the variation of the sum and difference squeezing parameters $V(a_1,a_2)$ and  $W(a_1,a_2)$, respectively. The negative values of the parameters $V(a_1,a_2)$/$W(a_1,a_2)$,for any $\phi$ confirm the existence of the sum/difference squeezing. It can be seen that the sum/difference squeezing is enhanced in PTS regime ($g/\gamma > 1$) as compared to PTSB regime ($g/\gamma < 1$).

\section{Conclusion}\label{conclusion}
 We considered a two cavity gain-loss system and discussed the conditions necessary for exhibiting PT invariance. This demanded  equal gain-loss in the two cavities. Further,  complete PT invariance requires the eigenvalues of the effective Hamiltonian to be real. This condition in turn means that the dominance of the gain/loss over the coupling strength $g$ breaks the PT invariance. With this setting, we studied the average photon number with different initial states, viz., vacuum, NOON, coherent, and thermal states. In all the four cases, the average photon number shows a similar behavior for gain and loss modes in PTS regime. In contrast to this, in PTSB regime, the gain mode is found to  dominate over the  lossy mode, while both show an exponential growth. We further studied some nonclassical features using the average photon numbers for different initial states. Specifically, we have reported the presence of QZE and QAZE in two cavities and 
nonclassical features, like intermodal antibunching and sum and difference squeezing. These witnesses of nonclassicality as well as the Zeno parameter exhibit suppression in the nonclassical features when one goes from PTS to PTSB regime. In other words, the dominance of the  loss/gain over the coupling strength results in  depletion of the  nonclassical features of the fields. Further,  it's observed that the relative phase of the input coherent fields provides us a control parameter to switch between QZE and QAZE.

The present study is expected to impact deeper understanding of PT symmetry and the role it can play to probe nonclassicality in the physical systems relevant in the field of quantum optics and information processing.
\section*{Acknowledgments}
The work of S.B. is supported by Project No. 03(1369)/16/EMR-II, funded by the Council of Scientific and Industrial Research, New Delhi. AP  thanks  Department  of  Science and  Technology (DST), India  for  the support provided through the project number EMR/2015/000393. KT thanks the project LO1305 of the Ministry of Education, Youth and Sports of the Czech Republic for support. Authors also thank Nasir Alam for some fruitful discussions.

\bibliographystyle{apsrev4-1}
\bibliography{PT_Symmetry}

\section*{Appendix A: }
\label{appendix:A}
\setcounter{equation}{0} \renewcommand{\theequation}{A.\arabic{equation}} 

\textit{Average photon number with initial state as a NOON state.} For a general NOON state $\frac{\ket{n,0} + \ket{0,n}}{\sqrt{2}}$, the average photon numbers can be shown to be
	
\begin{equation}
\begin{array}{lcl}
\langle a_1^\dagger (t) a_1(t) \rangle &=& \frac{1}{2 \Omega^3} \left\{-2\gamma (\gamma +g^2 t) \Omega + n \Omega^3 \cosh^2(\Omega t) \right.\\
&+& 2 \gamma^2 \Omega \cosh(2 \Omega t) + (g^2 + \gamma^2) n \Omega \sinh^2(\Omega t)\\
& +& \left.(n \Omega^2 + 2 \gamma^2 -g^2) \gamma \sinh(2 \Omega t)\right\},\\
\langle a_2^\dagger (t) a_2(t) \rangle &=& \frac{1}{2 \Omega^3}\left\{-2 g^2 \gamma \Omega t + n \Omega^3 \cosh^2 (\Omega t) \right. \\
&+& n \Omega(g^2 + \gamma^2)\sinh^2(\Omega t)  + \gamma (g^2 - n \Omega^2)\\
&\times& \left. \sinh(2\Omega t)\right\}.
\end{array}
\end{equation}

\textit{Average photon number with initial state as a coherent state.} With a coherent state of the form $\ket{\alpha_1, \alpha_2}$, such that $\alpha_1 = r_1e^{i\theta_1}$ and $\alpha_2 = r_2e^{i\theta_2}$, we have the following expression for the average photon numbers:

\begin{equation}
\begin{array}{lcl}
\langle a^\dagger_1(t) a_1(t) \rangle &=&  \frac{1}{2\Omega^3} \Bigg[ \gamma (-g^2 + 2\gamma^2) \sinh(2\Omega t) \\
&-&  2 \Omega\gamma (\gamma + g^2 t) +  2 \Omega r_1^2 \Omega^2 \cosh^2(\Omega t) \\
&+&  2 \Omega \gamma^2 \cosh(2\Omega t) +2 \Omega \sinh^2(\Omega t) \\
&\times& \bigg( \gamma^2 r_1^2 + g^2 r_2^2 + 2 g \gamma r_1 r_2 \sin(\Delta \theta) \bigg)\\
& +&  2 \Omega r_1  \bigg(\gamma r_1 + g r_2 \sin(\Delta \theta) \bigg) \sinh(2 \Omega t) \Bigg], \\
\langle a^\dagger_2(t) a_2(t) \rangle &=& r_2^2 \cosh^2(\Omega t) - \frac{r_2 \bigg( \gamma r_2 + g r_1 \sin(\Delta \theta) \bigg) \sinh(2\Omega t)}{\Omega} \\
&+& \frac{2 \Omega \bigg( g^2 r_1^2 + \gamma^2 r_2^2 + 2 g \gamma r_1 r_2 \sin(\Delta \theta ) \bigg) \sinh^2(\Omega t) }{2 \Omega^3}\\
&+& \frac{g^2 \gamma (-2 \Omega t + \sinh(2\Omega t))}{2 \Omega^3}.
\end{array}
\end{equation}

\textit{Average photon number with initial state as a thermal state.} The two mode isotropic thermal state can be represented by the normalized density matrix
\begin{align}
\rho_0(\beta) &= (1-e^\beta)^2 \exp[-\beta(a_1^\dagger a_1 + a_2^\dagger a_2)],\nonumber \\
       &= (1-e^\beta)^2 \sum_{n_1, n_2 = 0}^{\infty} \exp(-\beta (n_1+n_2)) |n_1,n_2 \rangle \langle n_1,n_2 |.
\end{align}
Here $\beta = \hbar \omega/ k_B T$ and we have used the natural units $\hbar = k_B =1$. The average photon number in this case is given by

\begin{equation}
\begin{array}{lcl}
	\langle a^\dagger_1(t) a_1(t) \rangle &=& \frac{g^2 \sinh^2(\Omega t)}{\Omega^2} (1-e^{\beta})^2 \sum_{n_1, n_2}^{} e^{-\beta (n_1 + n_2)} n_2 \\
	&+& \left\{ \cosh(\Omega t) + \frac{\gamma}{\Omega} \sinh(\Omega t)\right\}^2 (1-e^{\beta})^2\\
	&\times& \sum_{n_1, n_2}^{} e^{-\beta (n_1 + n_2)} n_1  -\frac{\gamma  (g^2 - 2\gamma^2) \sinh(2\Omega t) }{2 \Omega^3}\\
	&-& \frac{\gamma \left\{ 2 (\gamma + g^2 t) \Omega  - 2\gamma \Omega \cosh(2\Omega t)\right\}}{2 \Omega^3} \\
	\langle a^\dagger_2(t) a_2(t) \rangle &= &\frac{g^2 \sinh^2(\Omega t)}{\Omega^2} (1-e^{\beta})^2 \sum_{n_1, n_2}^{} e^{-\beta (n_1 + n_2)} n_2 \\
	&+&  \left\{ \cosh(\Omega t) + \frac{\gamma}{\Omega} \sinh(\Omega t)\right\}^2 \\
	&+& \frac{g^2 \gamma}{2 \Omega^3} (-2\Omega t + \sinh(2 \Omega t)).
	\end{array}
\end{equation}

\end{document}